\documentclass[journal]{IEEEtran}
\usepackage{blindtext}
\usepackage{graphicx}
\usepackage{color}
\usepackage{gensymb}
\usepackage{amsmath}
\usepackage{amssymb}

\begin{document}
\title{An Experimental Study on Microservices based Edge Computing Platforms}

\author{
\IEEEauthorblockN{Qian Qu, Ronghua Xu, Seyed Yahya Nikouei, Yu Chen}

\IEEEauthorblockA{Dept. of Electrical \& Computer Engineering, Binghamton University, SUNY,  Binghamton, NY 13902, USA \\
\{qqu2, rxu22, snikoue1, ychen\}@binghamton.edu}
}

\maketitle
\begin{abstract}
The rapid technological advances in the Internet of Things (IoT) allows the blueprint of Smart Cities to become feasible by integrating heterogeneous cloud/fog/edge computing paradigms to collaboratively provide variant smart services in our cities and communities. Thanks to attractive features like fine granularity and loose coupling, the microservices architecture has been proposed to provide scalable and extensible services in large scale distributed IoT systems. Recent studies have evaluated and analyzed the performance interference between microservices based on scenarios on the cloud computing environment. However, they are not holistic for IoT applications given the restriction of the edge device like computation consumption and network capacity. This paper investigates multiple microservice deployment policies on edge computing platform. The microservices are developed as docker containers, and comprehensive experimental results demonstrate the performance and interference of microservices running on benchmark scenarios.   
\end{abstract}




\section{Introduction}
\label{sec:introduct}
While cloud computing has changed human's life by providing service oriented architecture (SOA) as a service (SaaS), Platform as a Service (PaaS), Infrastructure as a Service (IaaS) \cite{fox2009above}, the proliferation of the Internet of Things (IoT), Artificial Intelligence (AI) and edge computing technologies leads us enter the era of post-cloud computing \cite{shi2016promise}. According to the estimation of Cisco Global Cloud Index, the total amount of data created by any device will reach 847 Zettabytes (ZB) annually by 2021 while the Internet traffic will reach 19.5 ZB by then \cite{index2017forecast}. Some of the produced data might require timeliness, involve privacy or cause unpredictable impact on the network. For example, in applications like smart traffic lights, real-time urban surveillance, cloud computing is not able to serve the purpose well \cite{chen2018smart, chen2017smart}.

Rapid technological advances in cloud computing and Internet of Things (IoT) make Smart Cities feasible by integrating heterogeneous computing devices to collaboratively provide variant pervasively deployed smart services \cite{nikouei2019kerman, nikouei2018smart}, in which the capability of data collection and processing at the edge is the key \cite{nikouei2019toward}. Migrating lower-level data processing tasks to edge devices enables the system to meet the requirements for delay-sensitive, mission-critical applications, including smart urban surveillance \cite{chen2017enabling}, instant privacy protection \cite{fitwi2019lightweight, fitwi2019no}, real-time public safety \cite{xu2018real}, etc. However, the heterogeneity and resource constraint at edge necessitate a scalable, flexible and lightweight system architecture that supports fast development and easy deployment among multiple service providers.

Because of many attractive features, such as good scalability, fine granularity, loose coupling, continuous development, and low maintenance cost, the microservices architecture has emerged and gained a lot of interests both in industry and academic community \cite{nagothu2018microservice, nikouei2019decentralized}. Compared to traditional SOAs in which the system is a monolithic unit, the microservices architecture divides an monolithic application into multiple atomic microservices that run independently on distributed computing platforms. Each microservice performs one specific sub-task or service, which requires less computation resource and reduces the communication overhead. Such characteristics make the microservices architecture an ideal candidate to build a flexible platform, which is easy to be developed and maintained for cross-domain applications like smart public safety, smart traffic surveillance systems, etc \cite{chen2018smart, nikouei2018smart}.

Motivated by the observations obtained in our previous research \cite{nikouei2019decentralized, xu2019blendmas}, this work tries to answer three questions:
\begin{itemize}
    \item Is it suitable to run multiple microservices inside one container just considering the performance of the edge device? Which type of microservices with different resource consuming inclination could be put together if the answer is yes?
    \item Is the effect of interference that executes multiple microservices running on fog computing and edge computing scenarios different? and
    \item What is the trade-off between the ``one process per container'' rule and the existing limitation of resource at edge side, such as computation and memory? 
\end{itemize}

The rest of this paper is organized as follows. Section \ref{sec:background} discusses the current arts of research on microservices technology and performance evaluation. The microservices deployment policy and performance matrix are explained in Section \ref{sec:systemdesign}. Given comprehensive experimental results, the performance evaluations are discussed in Section \ref{sec:experiment}. Finally, a summary is presented in Section \ref{sec:conclusions}. 

\section{Related Work}
\label{sec:background}

\subsection{Microservices Architecture}

Microservices architectures (MSA) are extensions of a Service Oriented Architecture (SOA) \cite{wilde2016approaches}. The traditional SOA uses a monolithic architecture that constitutes different software features in a single interconnected application and database. Relaying on the tightly coupled dependence among functions and components, it is more challenging to adapt to new requirements in an IoT-enabled system than the SOA, which requires scalability, service extensibility, and cross-platform interoperability \cite{datta2018next}. The main difference between MSA and SOA is that a MSA is more decentralized and it distributes all the logic (such as routing, message parsing) in to smart end points \footnote{http://www.opengroup.org/soa/source-book/msawp/p3.htm}. MSA also adopts a light weight applications programming interface (API) gateway for managing services instead of using heavier and more sophisticated Enterprise Service Bus \cite{lander2019microservices}. Each microservice runs its own process and communicates with peers using light weight communication mechanisms (REST API or SOAP protocol \cite{malavalli2015scalable}, etc). The services are built around business capabilities and independently deployed by fully automated deployment tools \cite{pathania2017setting}.

Fine granularity and loose coupling are two significant features in MSA \cite{yu2018survey}. Fine granularity means that there is a bare minimum of centralized management of these services. Moreover, there are instances where the service is not micro \cite{sheng2019micro}. Loose coupling implies that each of microservices components has few dependencies on other separate components, which makes the deployment and development of Micro-services more independent \cite{pautasso2017microservices}.

Granularity should be considered from the standpoint of its eventual impact on service performance as the decision process is influenced by numerous environmental factors when deploying MSA\cite{shadija2017microservices}.

\subsection{Performance Evaluation}

Container technology offers a more lightweight method to abstract the applications from the system environment which allows the microservices to be deployed quickly but also consistently. Compared to VMs, containers not only provide all the libraries and other dependencies but also consume less resource and produce lower overhead \cite{wu2015pseudo, wu2017container}. Applications with different dependencies are wrapped into a container image and shared to various users. These features allows the applications to be run at the edge of the network. Ruchika et al. \cite{tech2016evaluation} evaluated the feasibility of using container technology like docker for IoT applications.

Recent studies compared the performance of containers to virtual machines and showed that in most occasions containers work better than or almost equal to virtual machines. The IBM Research Division measured the performance of Docker in terms of CPU, memory, disk I/O and compared the result with KVM \cite{felter2015updated}. The research showed that in all evaluated cases container works better than VMs. In a study researchers have compared the performance of Docker with VMs when running the Spark applications  \cite{bhimani2017accelerating}. They claimed that Spark works better with docker for calculation intensive applications. Similar work has been done in big data area with the interference caused by neighbor containers running big data microservices \cite{ye2017performance}. The performance of different single-board computer devices are evaluated by deploying containerized benchmarks \cite{morabito2017virtualization}, which focused on hardware comparison and the choice of one device rather than another, however, it did not pay much attention on container analysis. An optimized version of Docker called Docker-pi was proposed for container deployment on single-board machines to speed up docker deployment \cite{ahmed2018docker}.

Researchers have studied the performance of collocated microservices running on a common host and compared the result in container with VMs \cite{sharma2016containers}. Evaluation is also reported on how microservices are interfered with each other in cloud computing environment \cite{jha2018holistic}, in which the performances between processed in a container and the situation where they follow the ``one process per-container'' rule are compared. The conclusion was that putting microservices inside one container has a positive effect to the overall performance. However, the work is based on a cloud computing platform.

A survey of VM management lists the virtualization frameworks for edge computing and elaborates the techniques tailored for it \cite{tao2019survey}. Focusing on Docker, a container based technology as an edge computing platform, researchers evaluated four criteria and conclude that Docker is a viable candidate \cite{ismail2015evaluation}. 

To the best of our knowledge, there is not a reported effort that evaluates the interference effect of containers in the edge computing environment. The trade-off between lower overheads and the ``one process per container'' rule seems more important considering the resource limitation at the edge.

\section{Microservices Benchmark Evaluation}
\label{sec:systemdesign}
For microservices development, we choose Docker container owing to popularity and familiarity in container community. Docker wraps up the application with all its dependencies into a container so that it can easily be executed on any Linux server (on-premise, bare metal, public or private cloud). It uses layered file system images along with the other Linux kernel features (namespace and cgroups) for the management of containers. This feature allows Docker to create any number of containers from a base image, which are copies of the base image wrapped up with additional features. This also reduces the overall memory and storage requirements that eases fast startup.

To measure the performance of the containerized microservices deployment policy, several widely used benchmarks are considered in designing the microservices benchmark based on typical computing paradigms used in the smart public safety system at the edge. The key metrics of the benchmark are defined as following:

\begin{enumerate}
    \item CPU performance: To evaluate the performance of the CPU, we chose the Linpack benchmarks which measure the system's floating point computing power by solving linear equations. And the capability of the CPU is measured in terms of FLOPS (floating point operations per second).
    \item Memory performance:We chose the STREAM benchmarks to measure the performance of the memory. STREAM is a simple synthetic benchmark program that measures sustainable memory bandwidth (in MB/s) and the corresponding computation rate for simple vector kernels.
    \item Disk performance: To measure the disk I/O capability, we chose the Bonnie++ benchmark which gives the results in terms of input, output, rewrite (in Kb/s) and seeks (in per second).
\end{enumerate}

\begin{figure}[b]
    \centering
    \includegraphics[height=0.42\textwidth]{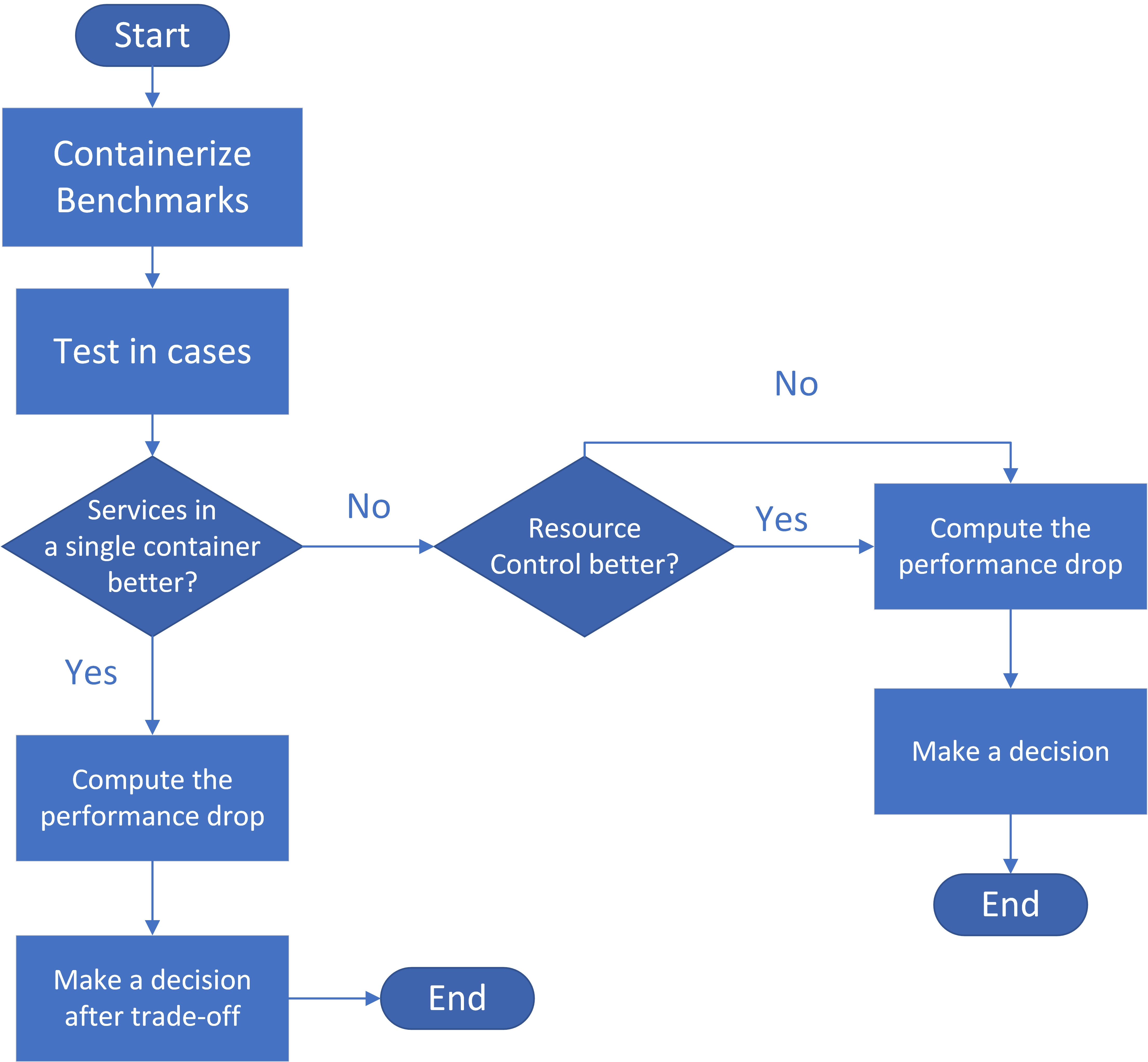}
    \caption{Containerized benchmark evaluation workflow.}
    \label{fig:workflow}
\end{figure}

For each metric evaluation, the containerized microservices are deployed both on fog (desktop) side and edge (Raspberry Pi 4) side. Given above defined benchmarks, the performance and the interference effect of the containerized microservices are evaluated based on a set of microservice deployment policies. Figure \ref{fig:workflow} illustrates the whole containerized benchmark evaluation workflow, and the microservice deployment policies are described as following four cases:

\begin{itemize}
    \item \emph{Case 1}: One microservice is developed as single container and only one container is running on host machine. Since we would put multiple microservices in one container later, the system resource is constrained by using the control groups or Cgroups, a Linux kernel feature. Such case is considered as a baseline for the entire evaluation.
    \item \emph{Case 2}: Multiple microservices are developed as single container and only one container is running on host machine. In this case we do not employ Cgroups so the microservices can have access to all the system resources. This case is used to evaluate resource competition caused by wrapping multiple microservices into one container. 
    \item \emph{Case 3}: Each container only holds single microservice and multiple containers are running on host machine without employing Cgroups. This case could evaluate resource competition caused by multiple containers deployed on single host platform.
    \item \emph{Case 4}: Each container only holds single microservice and multiple containers are running on host machine. Like case 1, this case employs Cgroups to limit the system resource during test.
\end{itemize}

\section{Experimental Analysis}
\label{sec:experiment}
\subsection{Experimental Setup}
A concept-proof prototype system has been implemented on a simulating SPS scenarios. Table \ref{tab:testbed} describes configurations of experimental testbed. The desk top simulates fog node while Raspberry Pi 4 acts as edge node.

\begin{table}[ht]
\caption{Configuration of Experimental Testbed.} 
\label{tab:testbed}
\begin{center}       
\begin{tabular}{|l|p{2.5cm}|p{3.5cm}|} 
\hline
\rule[-1ex]{0pt}{3.5ex} \textbf{Device} & Desktop & Raspberry Pi 4 \\
\hline
\rule[-1ex]{0pt}{3.5ex} \textbf{CPU} & 4.5GHz Intel Core TM (4 cores) & Quad core Cortex-A72 (ARM v8), 1.5GHz \\
\hline
\rule[-1ex]{0pt}{3.5ex} \textbf{Memory} & 4GB DDR3 & 4GB LPDDR4 RAM \\
\hline
\rule[-1ex]{0pt}{3.5ex} \textbf{Storage} & 50G HHD & 32GB (microSD card) \\
\hline
\rule[-1ex]{0pt}{3.5ex} \textbf{OS} & Ubuntu 16.04 & Raspbian GNU/Linux (Jessie) \\
\hline
\end{tabular}
\end{center}
\end{table}
\vspace{-10pt}

For workload size and configuration on fog, we provide a specific configuration for each microservice. For Linpack, we consider the matrix of size N as 15000. We also considered the problem size (number of equations to solve) as 15000. We configure the STREAM by setting the array size as 60M and DNTIMES as 200. The total memory requirement for this configuration is 1373.3 MiB. For Bonnie++, we considered the file size as 8192 MB and set the uid to use as root.

For workload size and configuration on edge (Raspberry Pi 4) side, each microservice, we provide a specific configuration. For Linpack, we consider the matrix of size N as 15000. We also considered the problem size (number of equations to solve) as 15000. We configure the STREAM by setting the array size as 30M and DNTIMES as 100. The total memory requirement for this configuration is 1373.3 MiB. For Bonnie++, we considered the file size as 4096 MB and set the uid to use as root.

The experimental test is conducted based on performance of benchmarks and microservices deployment policies. For the ease of presentation of the results, we used the following abbreviations for the microservices used in the simulation of SPS system, they are namely Linpack (L), STREAM (S), and Bonnie++ (B). To get experimental results, for case 1, each microservice (L, S and B) is performed 30 iterations. For case 2, case 3 and case 4, since the microservices are in all possible combinations and corresponding isolated running time of the containers are not identical, it is not suitable to repeat the process for a certain number of iterations. Hence, for these cases, we keep the container running for a certain period of time (60 minutes) and compute the average performance.

\subsection{Evaluation on Fog Side}

In this paper we define a desktop as the fog side, the most usual physical machine in a LAN system and  the role of a manager of the other edge computing device in our IoT system.

\begin{figure}[b]
    \centering
    \includegraphics[width=.45\textwidth]{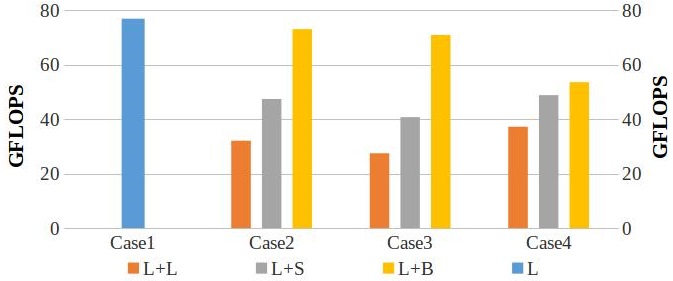}
    \caption{CPU Performance on Fog.}
    \label{fig:cpu_fog}
\end{figure}

\emph{1. CPU Performance and Analysis}:
According to the CPU architecture, we chose the Linpack version for Intel and implemented it in Docker. We ran the container(s) and recorded the performance of Linpack microservice as described in Section \ref{sec:systemdesign}. Figure \ref{fig:cpu_fog} shows the average CPU performance of running Linpack in terms of GFLOPS. Compared with case 1, all combination of microservices have more computation overhead on host performance. The L+B in case 2 has the least impact on CPU performance while L+L in case 3 is the worst result that has 64\% degradation compared to the baseline L in case 1. Owing to higher computing operations, like float calculation, used by Linpack (L), so L+L combination shows worst CPU performance in all deployment policies. Furthermore, in case 3, control groups (Cgroups) are disabled so that containers have to compete for the system resource. Therefore, test results in case 4 have better performance than those in case 3.

\begin{figure}[t]
    \centering
    \includegraphics[width=.45\textwidth]{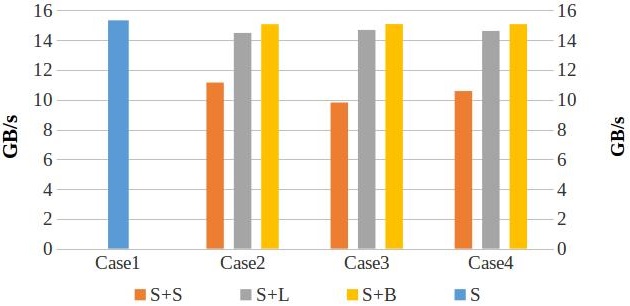}
    \caption{Bandwidth of COPY on Fog.}
    \label{fig:mem_fog}
    \vspace{-10pt}
\end{figure}

\noindent \emph{2. Memory Performance and Analysis}:
To evaluate the memory performance, we wrapped the STREAM benchmark inside a Docker container. The STREAM benchmark employs four vector operations and gives a result of bandwidth. The average result of all the four vector operations (COPY, SCALE, ADD and TRIAD) demonstrate similar shapes in bar chart, so we just present the result of COPY operation as memory performance. As shown in Fig. \ref{fig:mem_fog}, except for combination S+S in case 2, the others have limited performance deduction compared to the baseline in case 1. Given the results, wrapping memory type workloads microservices as one container introduce higher memory overhead on fog computing environment. Similar to the results in CPU performance, all the combinations show best performance in case 4 when Cgroups is enabled.

\begin{figure}[b]
    \centering
    \includegraphics[width=.48\textwidth]{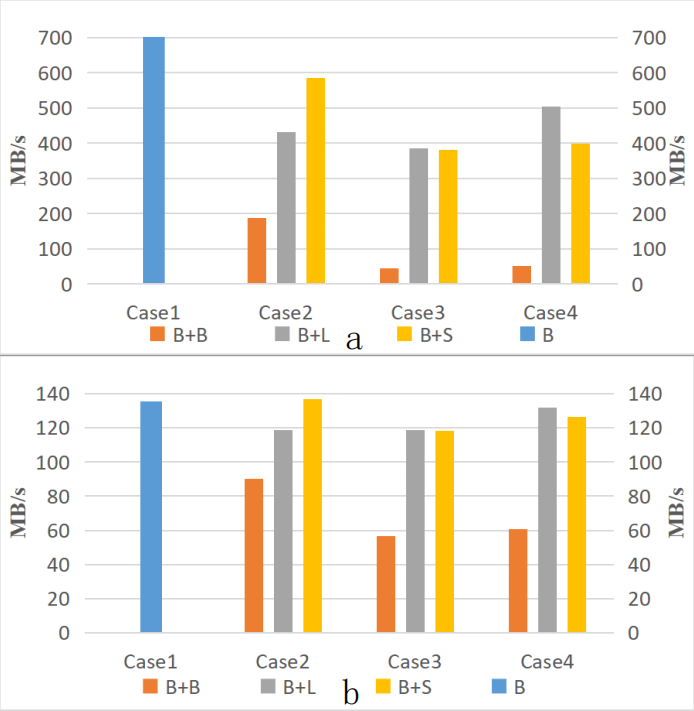}
    \caption{a) Performance of Input on Fog; b) Performance of Output on Fog.}
    \label{fig:IO_fog}
\end{figure}

\noindent \emph{3. Disk I/O Performance and Analysis}:
The Bonnie++ benchmark evaluates various performance parameters in which only sequential block input and output scenarios are considered during test. The performance evaluation is based on I/O throughput in terms of MB/s. Figure \ref{fig:IO_fog} shows that the input performance is better than output performance on fog node. The results show the similar interference pattern that running multiple instances of Bonnie++ leads to lower I/O throughput, no matter cases that are deployed in a single container or separate containers. From case 2 to case 4, the performance is nearly equal to the baseline case 1 except for the combination of B+B.

In our IoT system, for instance, a local smart surveillance, a fog device plays the role of managing and monitoring the performance of the edge device like smart cameras. Consider that the nodes could be numerous, the fog device is more sensitive to the granularity of the containers. Based on the evaluation results, generally it seems more practical to reduce the number of containers which are CPU consuming inclined. Meanwhile we should avoid deploying same type microservices together. The evaluation gives a general direction of setting up the fog device, we should adjust the deployment of the specified microservices according to the actual situation of the system and capability of the hardware.

\subsection{Evaluation on Edge Side}
The experimental process on edge ( Raspberry Pi) is quite the same as fog side. However, given limited resource on edge node, we set the time interval to 120 minutes for the following evaluations.

\begin{figure}[t]
    \centering
    \includegraphics[width=.45\textwidth]{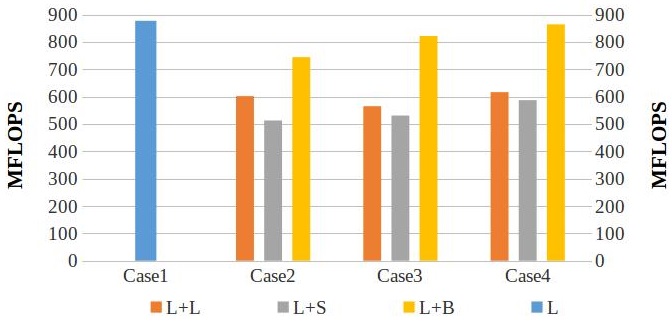}
    \caption{CPU Performance on Edge.}
    \label{fig:cup-edge}
    \vspace{-10pt}
\end{figure}

\noindent \emph{1. CPU Performance and Analysis}:
Since the CPU architecture of Pi 4 is ARM, we picked another Linpack version and adjusted the codes to the system environment. Then We containerized the benchmark and ran experiments as per the various case described in Section \ref{sec:systemdesign}. Figure \ref{fig:cup-edge} shows the CPU performance in terms of MFLOPS.

Similar to the results on fog side, all the combinations have a significant impact on the computing performance of host except for the combination of L+B. The combination of L+S in case 2 has worst performance, so it causes 41\% degradation compared to the baseline of case 1. But all three combinations demonstrate the best performance in case 4, knowing the fact that the microservices are separated in two containers with the control group enabled. This is quite different from the case in physical machine where L+B in case 4 did not give a good performance.

The results also indicate that computing type workloads tend to have the worst performance when working with memory consuming ones when they are wrapped into a single container with cgroups employed. And it shows that CPU consuming microservice works well with disk I/O workloads and the performance is much closer to the baseline relatively compared to the fog environment. In addition, as Fig. \ref{fig:cup-edge} shows, case 4 has better performance than case 3 while running all combinations. The reason behind is that in case 3 Cgroups are disabled so the containers have to compete for the system resource, however, the containers have no interference with each other in case 4 when Cgroups are enabled. 

\begin{figure}[t]
    \centering
    \includegraphics[width=.45\textwidth]{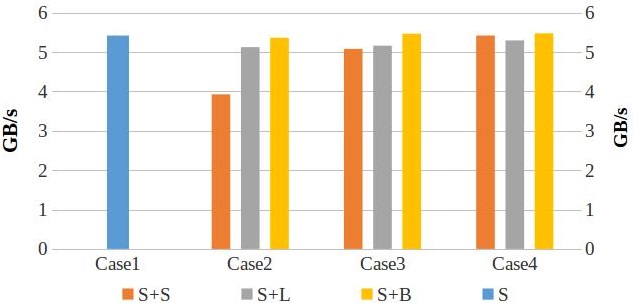}
    \caption{Bandwidth of COPY on Edge.}
    \label{fig:copy-edge}
    \vspace{-10pt}
\end{figure}

\noindent \emph{2. Memory Performance and Analysis}:
Similar to fog side, only the of COPY operation is used for evaluation for memory performance in terms of GB/s. As shown in Fig. \ref{fig:copy-edge}, all four vector operations generate almost the same shapes in charts. Compared to the baseline in case 1, the results shows that all combinations have limited performance reduction except for combination S+L in case 2. Hence, enclosing memory type workloads inside one container as microservice is not suggested on edge computing platform. Like the results in CPU performance, the combinations work best in case 4 that microservices are separated in containers with resource control. 

\begin{figure}[htbp]
    \centering
    \includegraphics[width=.48\textwidth]{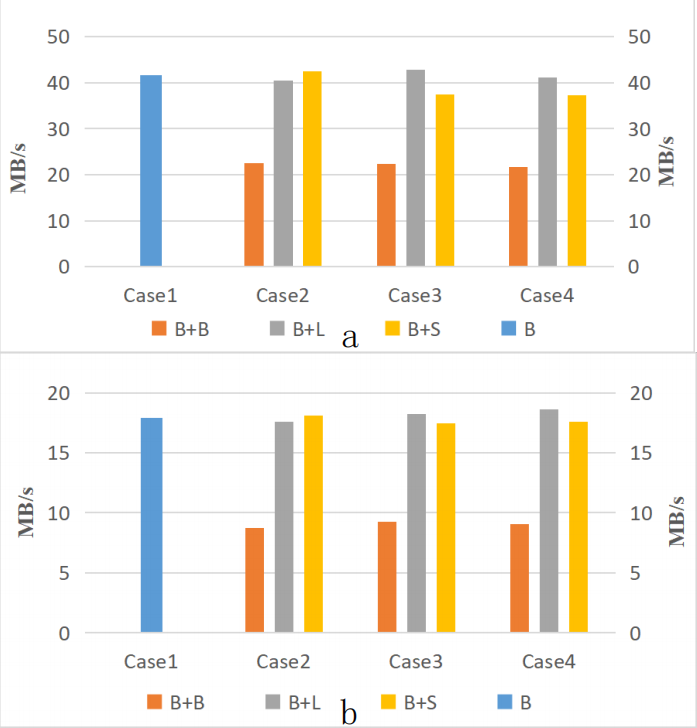}
    \caption{a) Performance of Input on Edge; b) Performance of Output on Edge.}
    \label{fig:IO-edge}
    \vspace{-10pt}
\end{figure}

\noindent \emph{3. Disk I/O Performance and Analysis}:
To evaluate Disk I/O performance on edge device, we used Bonnie++ microservice that creates a large file at least twice the size of inbuilt memory, and only sequential block input and output are considered during the test. The throughput performance of running microservices on edge is shown in Fig. \ref{fig:IO-edge}. The result shows the similar interference pattern indicating that running multiple instances of Bonnie++ creates higher resource contention when they are deployed either in a single container or separate containers. The results also indicate that the performance of combinations is nearly equal to the baseline performance except for the case of running multiple instances of Bonnie++ (B+B). All the combinations of deployment policies show very small deviation from the baseline of case 1 except for case of (B+B). 

Basically in our IoT system, the edge device like Raspberry Pi suffer from the limited resource so we should carefully consider the combination of microservices or simply isolate some inside a single container. Based on the results, generally we should deploy the CPU consuming inclined microservice alone, in smart surveillance system for example, the Computer Vision(CV) type containerized services. And it is considerable to wrap memory and disk inclined ones together to reduce the overheads and spare more system resources.

\section{Conclusions}
\label{sec:conclusions}
In this paper, a set of microservice deployment policies are proposed and performance matrix are considered. To evaluate the microservice deployment policy, a simulating test is developed and tested both on fog and edge computing platform. Given the experimental results, we conclude as following:

\begin{itemize}
    \item Under edge computing environment, running multiple microservices in a single container is an option considering the comparable performance (except for certain cases) if we ignore the ``\emph{one process per container}'' rule. The story behind the rule is that when deploy multiple process such as microservices in a single container we may face operations problem. For instance, we may have to replace the whole container if one of the microservices need updates which could be a disaster in a large scale. Considering the resource limitation, however, sometimes we have no choices.
    \item Memory intensive microservices are not heavily effected except for competing with same type ones inside one container. Disk I/O microservices is quite the same situation, however, both of the two kinds have a significant effect to the CPU intensive process.
\end{itemize}

This is a preliminary study toward a microservices architecture based smart public safety as an edge service. Our ongoing efforts include building a larger scale edge surveillance system with 32 smart cameras, which will enable a more comprehensive study for deeper insights.

\bibliographystyle{IEEEtranS}
\bibliography{sample-base}

\begin{thebibliography}{10}
\providecommand{\url}[1]{#1}
\csname url@samestyle\endcsname
\providecommand{\newblock}{\relax}
\providecommand{\bibinfo}[2]{#2}
\providecommand{\BIBentrySTDinterwordspacing}{\spaceskip=0pt\relax}
\providecommand{\BIBentryALTinterwordstretchfactor}{4}
\providecommand{\BIBentryALTinterwordspacing}{\spaceskip=\fontdimen2\font plus
\BIBentryALTinterwordstretchfactor\fontdimen3\font minus
  \fontdimen4\font\relax}
\providecommand{\BIBforeignlanguage}[2]{{%
\expandafter\ifx\csname l@#1\endcsname\relax
\typeout{** WARNING: IEEEtranS.bst: No hyphenation pattern has been}%
\typeout{** loaded for the language `#1'. Using the pattern for}%
\typeout{** the default language instead.}%
\else
\language=\csname l@#1\endcsname
\fi
#2}}
\providecommand{\BIBdecl}{\relax}
\BIBdecl

\bibitem{ahmed2018docker}
A.~Ahmed and G.~Pierre, ``Docker container deployment in fog computing
  infrastructures,'' in \emph{2018 IEEE International Conference on Edge
  Computing (EDGE)}.\hskip 1em plus 0.5em minus 0.4em\relax IEEE, 2018, pp.
  1--8.

\bibitem{bhimani2017accelerating}
J.~Bhimani, Z.~Yang, M.~Leeser, and N.~Mi, ``Accelerating big data applications
  using lightweight virtualization framework on enterprise cloud,'' in
  \emph{2017 IEEE High Performance Extreme Computing Conference (HPEC)}.\hskip
  1em plus 0.5em minus 0.4em\relax IEEE, 2017, pp. 1--7.

\bibitem{chen2018smart}
N.~Chen and Y.~Chen, ``Smart city surveillance at the network edge in the era
  of iot: opportunities and challenges,'' in \emph{Smart Cities}.\hskip 1em
  plus 0.5em minus 0.4em\relax Springer, 2018, pp. 153--176.

\bibitem{chen2017enabling}
N.~Chen, Y.~Chen, E.~Blasch, H.~Ling, Y.~You, and X.~Ye, ``Enabling smart urban
  surveillance at the edge,'' in \emph{2017 IEEE International Conference on
  Smart Cloud (SmartCloud)}.\hskip 1em plus 0.5em minus 0.4em\relax IEEE, 2017,
  pp. 109--119.

\bibitem{chen2017smart}
N.~Chen, Y.~Chen, X.~Ye, H.~Ling, S.~Song, and C.-T. Huang, ``Smart city
  surveillance in fog computing,'' in \emph{Advances in mobile cloud computing
  and big data in the 5G era}.\hskip 1em plus 0.5em minus 0.4em\relax Springer,
  2017, pp. 203--226.

\bibitem{datta2018next}
S.~K. Datta and C.~Bonnet, ``Next-generation, data centric and end-to-end iot
  architecture based on microservices,'' in \emph{2018 IEEE International
  Conference on Consumer Electronics-Asia (ICCE-Asia)}.\hskip 1em plus 0.5em
  minus 0.4em\relax IEEE, 2018, pp. 206--212.

\bibitem{felter2015updated}
W.~Felter, A.~Ferreira, R.~Rajamony, and J.~Rubio, ``An updated performance
  comparison of virtual machines and linux containers,'' in \emph{2015 IEEE
  international symposium on performance analysis of systems and software
  (ISPASS)}.\hskip 1em plus 0.5em minus 0.4em\relax IEEE, 2015, pp. 171--172.

\bibitem{fitwi2019lightweight}
A.~Fitwi, Y.~Chen, and S.~Zhu, ``A lightweight blockchain-based privacy
  protection for smart surveillance at the edge,'' in \emph{2019 IEEE
  International Conference on Blockchain (Blockchain)}.\hskip 1em plus 0.5em
  minus 0.4em\relax IEEE, 2019, pp. 552--555.

\bibitem{fitwi2019no}
------, ``No peeking through my windows: Conserving privacy in personal
  drones,'' \emph{arXiv preprint arXiv:1908.09935}, 2019.

\bibitem{fox2009above}
A.~Fox, R.~Griffith, A.~Joseph, R.~Katz, A.~Konwinski, G.~Lee, D.~Patterson,
  A.~Rabkin, and I.~Stoica, ``Above the clouds: A berkeley view of cloud
  computing,'' \emph{Dept. Electrical Eng. and Comput. Sciences, University of
  California, Berkeley, Rep. UCB/EECS}, vol.~28, no.~13, p. 2009, 2009.

\bibitem{index2017forecast}
C.~G.~C. Index and C.~C. V.~N. Index, ``Forecast and methodology, 2016--2021;
  white paper; cisco systems,'' \emph{Inc.: San Jose, CA, USA}, 2017.

\bibitem{ismail2015evaluation}
B.~I. Ismail, E.~M. Goortani, M.~B. Ab~Karim, W.~M. Tat, S.~Setapa, J.~Y. Luke,
  and O.~H. Hoe, ``Evaluation of docker as edge computing platform,'' in
  \emph{2015 IEEE Conference on Open Systems (ICOS)}.\hskip 1em plus 0.5em
  minus 0.4em\relax IEEE, 2015, pp. 130--135.

\bibitem{jha2018holistic}
D.~N. Jha, S.~Garg, P.~P. Jayaraman, R.~Buyya, Z.~Li, and R.~Ranjan, ``A
  holistic evaluation of docker containers for interfering microservices,'' in
  \emph{2018 IEEE International Conference on Services Computing (SCC)}.\hskip
  1em plus 0.5em minus 0.4em\relax IEEE, 2018, pp. 33--40.

\bibitem{lander2019microservices}
V.~Lander, D.~CARRU, A.~Sondhi, G.~Wilson \emph{et~al.}, ``Microservices based
  multi-tenant identity and data security management cloud service,'' Feb.~5
  2019, uS Patent App. 10/200,358.

\bibitem{malavalli2015scalable}
D.~Malavalli and S.~Sathappan, ``Scalable microservice based architecture for
  enabling dmtf profiles,'' in \emph{2015 11th International Conference on
  Network and Service Management (CNSM)}.\hskip 1em plus 0.5em minus
  0.4em\relax IEEE, 2015, pp. 428--432.

\bibitem{morabito2017virtualization}
R.~Morabito, ``Virtualization on internet of things edge devices with container
  technologies: A performance evaluation,'' \emph{IEEE Access}, vol.~5, pp.
  8835--8850, 2017.

\bibitem{nagothu2018microservice}
D.~Nagothu, R.~Xu, S.~Y. Nikouei, and Y.~Chen, ``A microservice-enabled
  architecture for smart surveillance using blockchain technology,'' in
  \emph{2018 IEEE International Smart Cities Conference (ISC2)}.\hskip 1em plus
  0.5em minus 0.4em\relax IEEE, 2018, pp. 1--4.

\bibitem{nikouei2019toward}
S.~Y. Nikouei, Y.~Chen, S.~Song, B.-Y. Choi, and T.~R. Faughnan, ``Toward
  intelligent surveillance as an edge network service (isense) using
  lightweight detection and tracking algorithms,'' \emph{IEEE Transactions on
  Services Computing}, 2019.

\bibitem{nikouei2019kerman}
S.~Y. Nikouei, Y.~Chen, S.~Song, and T.~R. Faughnan, ``Kerman: A hybrid
  lightweight tracking algorithm to enable smart surveillance as an edge
  service,'' in \emph{2019 16th IEEE Annual Consumer Communications \&
  Networking Conference (CCNC)}.\hskip 1em plus 0.5em minus 0.4em\relax IEEE,
  2019, pp. 1--6.

\bibitem{nikouei2018smart}
S.~Y. Nikouei, Y.~Chen, S.~Song, R.~Xu, B.-Y. Choi, and T.~Faughnan, ``Smart
  surveillance as an edge network service: From harr-cascade, svm to a
  lightweight cnn,'' in \emph{2018 ieee 4th international conference on
  collaboration and internet computing (cic)}.\hskip 1em plus 0.5em minus
  0.4em\relax IEEE, 2018, pp. 256--265.

\bibitem{nikouei2019decentralized}
S.~Y. Nikouei, R.~Xu, Y.~Chen, A.~Aved, and E.~Blasch, ``Decentralized smart
  surveillance through microservices platform,'' in \emph{Sensors and Systems
  for Space Applications XII}, vol. 11017.\hskip 1em plus 0.5em minus
  0.4em\relax International Society for Optics and Photonics, 2019, p. 110170K.

\bibitem{pathania2017setting}
N.~Pathania, ``Setting up jenkins on docker and cloud,'' in \emph{Pro
  Continuous Delivery}.\hskip 1em plus 0.5em minus 0.4em\relax Springer, 2017,
  pp. 115--143.

\bibitem{pautasso2017microservices}
C.~Pautasso, O.~Zimmermann, M.~Amundsen, J.~Lewis, and N.~M. Josuttis,
  ``Microservices in practice, part 1: Reality check and service design.''
  \emph{IEEE Software}, vol.~34, no.~1, pp. 91--98, 2017.

\bibitem{shadija2017microservices}
D.~Shadija, M.~Rezai, and R.~Hill, ``Microservices: granularity vs.
  performance,'' in \emph{Companion Proceedings of the10th International
  Conference on Utility and Cloud Computing}, 2017, pp. 215--220.

\bibitem{sharma2016containers}
P.~Sharma, L.~Chaufournier, P.~Shenoy, and Y.~Tay, ``Containers and virtual
  machines at scale: A comparative study,'' in \emph{Proceedings of the 17th
  International Middleware Conference}.\hskip 1em plus 0.5em minus 0.4em\relax
  ACM, 2016, p.~1.

\bibitem{sheng2019micro}
X.~Sheng, S.~Hu, and Y.~Lu, ``The micro-service architecture design research of
  financial trading system based on domain engineering,'' in \emph{2018
  International Symposium on Social Science and Management Innovation (SSMI
  2018)}.\hskip 1em plus 0.5em minus 0.4em\relax Atlantis Press, 2019.

\bibitem{shi2016promise}
W.~Shi and S.~Dustdar, ``The promise of edge computing,'' \emph{Computer},
  vol.~49, no.~5, pp. 78--81, 2016.

\bibitem{tao2019survey}
Z.~Tao, Q.~Xia, Z.~Hao, C.~Li, L.~Ma, S.~Yi, and Q.~Li, ``A survey of virtual
  machine management in edge computing,'' \emph{Proceedings of the IEEE}, 2019.

\bibitem{tech2016evaluation}
M.~Tech, ``Evaluation of docker for iot application,'' \emph{International
  Journal on Recent and Innovation Trends in Computing and Communication},
  vol.~4, no.~6, pp. 624--628, 2016.

\bibitem{wilde2016approaches}
N.~Wilde, B.~Gonen, E.~El-Sheikh, and A.~Zimmermann, ``Approaches to the
  evolution of soa systems,'' in \emph{Emerging Trends in the Evolution of
  Service-Oriented and Enterprise Architectures}.\hskip 1em plus 0.5em minus
  0.4em\relax Springer, 2016, pp. 5--21.

\bibitem{wu2015pseudo}
R.~Wu, B.~Liu, Y.~Chen, E.~Blasch, H.~Ling, and G.~Chen, ``Pseudo-real-time
  wide area motion imagery (wami) processing for dynamic feature detection,''
  in \emph{2015 18th International Conference on Information Fusion
  (Fusion)}.\hskip 1em plus 0.5em minus 0.4em\relax IEEE, 2015, pp. 1962--1969.

\bibitem{wu2017container}
------, ``A container-based elastic cloud architecture for pseudo real-time
  exploitation of wide area motion imagery (wami) stream,'' \emph{Journal of
  Signal Processing Systems}, vol.~88, no.~2, pp. 219--231, 2017.

\bibitem{xu2019blendmas}
R.~Xu, S.~Y. Nikouei, Y.~Chen, E.~Blasch, and A.~Aved, ``Blendmas: A
  blockchain-enabled decentralized microservices architecture for smart public
  safety,'' \emph{arXiv preprint arXiv:1902.10567}, 2019.

\bibitem{xu2018real}
R.~Xu, S.~Y. Nikouei, Y.~Chen, A.~Polunchenko, S.~Song, C.~Deng, and T.~R.
  Faughnan, ``Real-time human objects tracking for smart surveillance at the
  edge,'' in \emph{2018 IEEE International Conference on Communications
  (ICC)}.\hskip 1em plus 0.5em minus 0.4em\relax IEEE, 2018, pp. 1--6.

\bibitem{ye2017performance}
K.~Ye and Y.~Ji, ``Performance tuning and modeling for big data applications in
  docker containers,'' in \emph{2017 International Conference on Networking,
  Architecture, and Storage (NAS)}.\hskip 1em plus 0.5em minus 0.4em\relax
  IEEE, 2017, pp. 1--6.

\bibitem{yu2018survey}
D.~Yu, Y.~Jin, Y.~Zhang, and X.~Zheng, ``A survey on security issues in
  services communication of microservices-enabled fog applications,''
  \emph{Concurrency and Computation: Practice and Experience}, p. e4436, 2018.

\end{thebibliography}

\end{document}